\documentclass[preprint2]{aastex}

\begin{document}

\title{Driving spiral arms in the circumstellar disks of HD 100546 and HD 141569A}

\author{
Alice C.~Quillen, Peggy Varni\`ere, Ivan Minchev, \& Adam Frank
}
\affil{Department of Physics and Astronomy,
University of Rochester, Rochester, NY 14627}
\email{aquillen, pvarni, iminchev, afrank@pas.rochester.edu}

\begin{abstract}
With 2D hydrodynamical simulations of disks perturbed
externally by stars, brown dwarfs or planets we investigate possible scenarios
that can account for the spiral structure in circumstellar disks.
%
We consider two scenarios, spiral structure driven by an external bound 
planet or low mass star and that excited by 
a previous stellar close encounter or flyby.
We find that both scenarios produce morphology similar to
that observed in the outer disks of HD 141569A and HD 100546; moderately
open 2-armed outer spiral structure.  
Our simulations exhibit some trends.
While bound prograde objects effectively truncate a disk, a close encounter
by a star instead pulls out spiral arms, spreading out
the disk.  Following a flyby the morphology 
of the spiral structure can be used to limit 
the mass of the perturbing object and the time since the encounter occurred.
Eccentric bound planetary perturbers tend to more efficiently clear gas
away from the planet, resulting in a reduction in the excited spiral amplitude.
Bound perturbers in thicker disks excite more open but lower amplitude
spiral structure.
When the bound object is higher mass (stellar), the disk is truncated
by the Roche lobe of the planet at periapse, and each time the companion
approaches periapse, spiral arms may be pulled out from the disk.
Thinner disks tend to exhibit more steeply truncated disk edges.

We find that the outer two-armed spiral structure at radii
greater than 300 AU observed in the disk of HD 141569A 
is qualitatively reproduced with tidal
perturbations from its companion binary HD 141569B,C on a prograde 
orbit near periapse.
Our simulation accounts for the outer spiral arms, but is less
successful than the secular model of Augereau and Papaloizou at matching
the lopsidedness or asymmetry of the disk edge at 300AU. 
Our simulations suggest that
the disk has been previously truncated by the tidal force from the binary
and has aspect ratio or thickness $h/r \lesssim 0.1$.   
The simulated disk also exhibits an enhanced density
at the disk edge.

We find that a bound object (stellar or planetary) is unlikely to explain 
the spiral structure in the disk of HD 100546.  
A co-eval planet or brown dwarf in the disk of sufficient mass to account for the amplitude 
of the spiral structure would be detectable
in NICMOS and STIS images, however existing images reveal no such object.
A previous encounter could explain the observed structure,
provided that the disk is thin, the mass of the perturbing star 
is greater than $\sim 0.1 M_\odot$, and the encounter occurred less than a few
thousand year ago.   This suggests that the object responsible
for causing the spiral structure is currently 
within a few arcminutes of the star.
However, the USNO-B proper motion survey reveals no candidate object 
associated with HD 100546
or any moving object that could have recently encountered HD 100546.
Furthermore, the probability that a field
star encountered HD 100546 in the past few thousand years
is very low.

\end{abstract}


\section{Introduction}


Recent high angular resolution studies of nearby young stars 
systems have revealed structure in the disks surrounding these stars.
For example, gaps and holes in disks 
have been observed via direct or coronagraph
imaging in systems such as HR 4796A and HD 141569A
\citep{jay,augereau99a,augereau99b,weinberger}, and intricate spiral
structure has been revealed in the circumstellar dusty disks surrounding
HD~100546 and HD~141569A \citep{grady,clampin}.
These two disks are very large (outer radius a few hundred AU) and 
the spiral structure is suspected to be caused by 
a perturber external to the system, such as a recent stellar encounter in 
the case of HD 100546 or its companion binary, HD 141569B and C, in the case
of HD 141569A \citep{grady,clampin,augereau2004}.

While HD 100546 is relatively isolated,  most stars are 
born in embedded stellar clusters \citep{lada}.    
HD 141569A is in a low density region near a molecular cloud and
has 3 nearby young stars in its vicinity \citep{weinberger2000}.
During the time that a star is part of its birth cluster, 
large circumstellar disks can be perturbed by 
close encounters with stars (e.g., see \citealt{adams}).
Many stars are part of binary systems.
Consequently the nearby stellar environment may be a key factor 
influencing the extent and evolution of circumstellar disks.
By studying the structure observed in the largest of these disks, 
we hope to better understand the
effect of external perturbers on these disks.
We can also probe for the dependence of the spiral structure on the 
properties of the disk itself, thus allowing us to learn about the formation
and evolution of extra-solar planetary systems.

HD 100546 is a nearby southern Herbig Be star
(KR Mus; B9.5~Ve; distance $d=103\pm 6$pc)
with an age estimated $t \sim 10$Myrs \citep{vandenancker}.
HD 141569A (B9.5~Ve) is similar with
$d=99 \pm 10$pc, age $t \sim 5$Myrs \citep{weinberger2000}.
Infrared spectroscopy with the Infrared Space Observatory 
revealed solid state emission features in HD 100546 similar to those seen
in the spectrum of comet Hale-Bopp \citep{malfait,bouwman}.
While many studies of circumstellar disks have been carried out with 
spectra \citep{Hu, malfait,li2003}, recent imaging studies have been able
to resolve the disks out to the large distance  
of 3--6" or 300--600AU 
\citep{augereau99b, augereau01, grady,pantin, weinberger2000,weinberger}.
The HST/STIS and ACS studies, in particular, have revealed the presence 
of intricate structure in these disks, including 2-armed spiral
structure at a radius of a few arcseconds 
or a few hundred AU from the star.  
The opening angle is large enough that spiral structure is clearly defined in
the STIS and ACS images. 
For HD 100546, the amplitude of
the spiral structure is probably low.  From the surface brightness 
profiles shown by \citet{grady} we estimate 
$A = {dS\over S}\sim 0.15$ where $S$ is the surface brightness
at optical wavelengths and $dS$ represents the size of azimuthal
variations in $S$.
Because of the difficulty removing the residual diffracted and
scattered light from the star (as discussed by \citealt{grady}),
this amplitude is uncertain.
For HD 141569A, the spiral structure is high amplitude/contrast in the outermost
regions, and lower amplitude at smaller radii \citep{clampin}.
We note that because of the different instrument characteristics and 
filter bandpasses, care should be used in comparing the STIS images 
(HD 100546; \citealt{grady}, HD141569A; \citealt{mouillet}) 
with ACS images  (HD 141569A; \citealt{clampin}).
While HD 141569A's disk edge is sharply defined, HD 100546
has an envelope that extends out 
to a radius larger than 1000 AU \citep{grady}.

As suggested by \citet{grady,clampin},
massive external perturbers could be responsible for 
exciting the observed spiral structure in these circumstellar disks.
Alternative possibilities include gravitational instability 
(e.g., \citealt{fukagawa04}), 
and an internal perturber such as a planet or brown dwarf 
(e.g., \citealt{mouillet}).
We first consider the possibility that the disk could
be gravitational unstable to the formation of spiral density waves.
The Toomre $Q$ parameter for a gaseous disk
$Q \sim {  c_s \kappa \over \pi G \Sigma}$,
where $\kappa$, the epicyclic frequency at radius $r$ is equivalent to the angular
rotation rate $\Omega$ for a Keplerian disk.  Here $\Sigma$ is the gas
density per unit area, and $c_s$ is the sound speed \citep{B+T}.
When $Q$ drops to near 1, the disk becomes unstable to
the growth of spiral density waves.
Hydrostatic equilibrium can be used to relate the disk
thickness to its sound speed,
${h\over r} \sim {c_s \over v}$, where
the disk aspect ratio is $h/r$,
$h$ is the vertical scale height of the disk and $v$
is the circular velocity.
Assuming a Keplerian disk, and approximating the mass of the disk
$M_d \sim \pi \Sigma r^2$,
we find $Q \sim { M_* \over M_d} {h \over r} $,
where $M_*$ is the mass of the star.
From this we see that if the disk has a low density,
a very thin and cold disk is required for
$Q$ to drop below one or two, allowing the growth
of spiral density waves in the disk.
Gas densities are expected to be low in circumstellar disks.
For example,
when ${M_* \over M_d} \ga 100$, only an extremely thin
disk, $h/r \la 0.01$
would be unstable, allowing $Q\lesssim 1$.  
Because this aspect ratio is extreme, we must
consider alternative possibilities to account for the spiral structure.

The dispersion relation for tightly
wound hydrodynamic waves in a two-dimensional
gaseous disk suggests that 
an internal perturber (such as a planet) would produce spiral arms that are more
tightly wound with increasing radius from the star (e.g., \citealt{pp3}). 
However, the spiral arms in the disks of HD 100546 and HD 141569A
are more open (less tightly wound) with increasing radius, 
suggesting that the force responsible is increasing with radius.
This would be true if the spiral structure was
caused by an external tidal force or caused by 
density waves excited from an external massive body.
Consequently we consider here massive perturbers that are located
exterior to the location of the spiral arms.

When a star has a close encounter with a planetesimal disk, the tidal
force from the star can cause an inclination change in the outer disk 
\citep{larwood},
scatter objects in an extra-solar Oort cloud analog 
\citep{kalas01},  and cause short lived spiral structure 
in a gaseous disk \citep{pfalzner,clarke,boffin,kory}.
%
Alternatively, in the context of stellar binaries, 
previous studies have shown that if a secondary star is exterior to a disk,
spiral density waves are driven into the disk \citep{artymowicz94,savonije}.
Most of the binary/disk simulations of \citet{savonije} 
showed prominent 2-armed structure
similar in morphology to the outermost parts of the HD~100546 system.
Tidal forces from a binary can also cause the outermost part of disk
to be lopsided, as shown in the case of HD 141569A \citep{augereau2004}.

In this paper we consider these two scenarios
that could account for the spiral structure in HD 100546 and HD 141569A.
We investigate the feasibility of these scenarios 
by carring out hydrodynamic simulations.  With these simulations we
explore the types of encounters can produce 2 armed spiral structure
similar to that observed in the disks of HD 100546 and HD 141569A.
We aim to produce numerical simulations that exhibit
two outer open spiral arms and roughly match 
the spiral amplitudes in the disks of HD 100546 and HD 141569A.
In Section 2 we describe our simulations.
In Section 3 we investigate the possibility that an object bound
to the star, external to the disk, could tidally
excite spiral structure and drive density waves into the disk.
In Section 4
we explore the possibility that a recent nearby stellar encounter 
causes the observed spiral structure.
In Section 5 we consider observational constraints on the 
possible external perturber responsible for the spiral structure
in the disk of HD 100546.
A summary and discussion follows.

\section{Simulation of spiral structure caused by external perturbers}

To determine the likely cause of the spiral structure in the outer parts of
these circumstellar disks, we have carried out two series of simulations.
Because we have little information about the vertical geometry and structure  
of these disks,
we have restricted this initial study to 2-dimensional hydrodynamic 
simulations, confining our study to the plane containing the disks.  

Both planetesimal and gaseous disks can be approximated as 
fluid disks that are described with a sound speed and a kinematic 
viscosity.  For planetesimal disks 
the sound speed depends on the particle velocity
dispersion, but for gaseous disks it is expected to depend on turbulent motions.
Based on the images by \citet{grady} we estimate that
fine features in HD 100546's disk have widths 
$\sim 0.3''$ at a radius from the star of $\sim 3''$. 
in HD 141569A's disk exhibits features with widths $\sim 0.6''$ at a radius
from the star of $\sim 4''$.
The width of these features suggests that structure exists in these disks
at a scale height $h$ with vertical aspect ratio $h/r \lesssim 0.15$.
However, because we may only be seeing an illuminated
surface (e.g., \citealt{dullemond}), 
these features may not be directly related to density variations
in the disk.
Nevertheless if the features observed are interpreted
to be in the disk, the disk is not likely to be extremely thick $h/r > 0.2$.
Consequently we assume here that the dynamics of these disks
may be approximated with 2-dimensional simulations.

Scattered light optical and near-infrared imaging 
and thermal infrared observations show that
the disks of HD 100546 and HD 141569A are dusty 
\citep{malfait,bouwman,li2003, augereau2004, weinberger}.
Both disks are strongly depleted of gas; HCO+ was
not detected in HD 100546, \citet{wilner03}, though 
cold CO has been detected at low levels in HD141569A \citet{zuckerman95}.
Gas has been detected in the inner few tens of AU in 
HD 100546 from UV absorption lines of atomic species and molecular
hydrogen \citep{deleuil,lecavelier}, and in
CO and H3+ from HD141569A \citet{brittain}.
The small dust grains detected in infrared emission from HD141569A 
need to be replenished on timescales shorter
than the age of the system \citet{li2003,augereau2004},
consequently HD 141569A is most likely a young debris disk.
The vertical optical depth of the HD 141569A's disk 
in visible wavelengths at 200 AU  is low, 
$\tau \sim 0.024$ \citep{li2003}.
HD 100546's disk intercepts about 50\% of the stellar light
and emits prominantly in the  mid infrared \citep{bouwman}.  
The low level of hot dust emitting in the near-infrared, 
and the presence of crystalline silicate emission features
suggest that the disk is depleted at around 10AU 
and flared outside this radius, and that planets may have formed
in the inner region \citep{bouwman}.
The disk is probably optically thick in visible wavelengths
at 10-30 AU but may not be at larger radii \citep{bouwman}.
There is no direct evidence based on the modeling of the infrared
spectrum that HD100546 is a debris (rather than gaseous) 
disk, though \citet{grady97}
suggested that planetesimals exist in the disk based on spectroscopic
variability that could be interpreted in terms of stargrazing comets.

Though the composition of these disks is not precisely known,
the possible presence of planetesimals in these disks suggests
that a collisionless particle simulations 
(e.g., \citealt{pfalzner,augereau2004})
might be a better match to their physical properties.  
Here we are carrying out hydrodynamic simulations rather
than collisionless particle simulations.
The primary advantage of the grid based hydrodynamic simulations is that
structure can be seen across a fairly large dynamic range in density.  This
allows us to probe structure at disk edges and resolve small differences
in density due to spiral structure.  To resolve low intensity spiral structure,
particularly at the edge of a disk (such as is seen in HD141569A) or
low amplitude structure (such as is seen in HD100546) a large
number of simulated collisionless particles would be required.
For younger systems such as HD100546, which has a high optical
depth, collisions could be frequent enough that particle motions
are damped. Due to shocks and viscous effects, hydrodynamic simulations
will damp non-circular motion.
While neither purely collisionless particle simulations or hydrodynamic
simulations are perfect physical models for debris disks, because
we can better resolve faint or low amplitude structure, 
our simulations are complementary to those carried out
using simulations comprised of collisionless particles
(e.g., \citealt{pfalzner,augereau2004}).

Both hydrodynamic and particle simulated disks support spiral structure,
however there are differences in the dispersion relation for spiral
density waves.  High frequency spiral density waves (similar
to sound waves) are carried by hydrodynamic
disks but are damped in particle disks.
In both cases the effective sound speed can be estimated using
an estimate for the disk thickness and vertical hydrostatic equilibrium.

Our simulations are carried out
using the hydrodynamical code
developed by \citet{masset02,masset03}. This code is an Eulerian
polar grid code with a staggered mesh and an artificial second
order viscous pressure to stabilize the shocks (see also
\citealt{stone}). The hydrocode allows tidal interaction between
one or more massive bodies (planets) and a 2D non-self-gravitating gaseous disk,
and is endowed with a fast advection algorithm that  removes the
average azimuthal velocity for the Courant timestep limit
\citep{masset00}.  The simulations are performed in the
non-inertial non-rotating frame centered on the primary star
(similar to a heliocentric frame).   
A logarithmic grid in radius was adopted.

The code is scaled so that the unit length is the outer edge of the disk,
the unit mass is that of the central star.
Time is given in units of one orbital period
for a particle in a circular orbit at the outer edge of the disk.
For both HD 100546 and HD 141569A with a disk of radius $\sim 300$AU and stellar mass
of $M_* \sim 2.5 M_\odot$, this period is $P_{outer} \sim 3300$ years.

The disk aspect ratio is defined as $h/r$ 
where $h$ is the vertical scale height of the disk.
The disk aspect ratio is assumed to be uniform and constant for each simulation.
The sound speed of the gas is set from the disk aspect ratio.
From hydrostatic equilibrium, we have $h/r \sim c_s/v_c$ where $v_c$ is
the velocity of an object in a circular orbit and $c_s$ is the sound speed
or velocity dispersion.
The simulations are begun with a disk with a smooth and azimuthally
symmetric density profile
$\Sigma(r) \propto 
\left({r \over R_{min}}\right)^{-1}$ where $R_{min}$
is the inner disk radius.
Because the tidal force from the perturber can pull gas outward,
we set the outer boundary to be open, allowing mass to flow outward.  
The grid inner boundary 
only allows gas to escape so that the disk material may be
accreted on to the central star.

Both particle and gaseous disks exhibit viscosity.
In gaseous disks the viscosity is usual related to the $\alpha$ parameter,
whereas in particle disks, an effective viscosity is dependent
on the opacity of the disk, (e.g., \citealt{M+D}).
In our simulations the disk viscosity is parametrized with the Reynolds number
${\cal R} \equiv r^2 \Omega/\nu$, where $r$ is the
radius, $\nu$ is the kinematic viscosity,
and $\Omega$ the Keplerian angular rotation rate of a particle
in a circular orbit of radius $r$.
In these simulations, 
the Reynolds number is assumed to be constant with radius.  
The Reynolds number determines the timescale in rotation
periods over which viscosity can cause accretion.  
The ages of these systems 5--10 million years
is only a thousand times the orbital period at the outer disk edges.
Because of the short number of orbital periods comprising the lifetime of
these disks, viscosity probably does not strongly affect
the recent dynamics.   We have chosen a Reynolds number
for our simulations (${\cal R} = 10^6$) which is high enough
that viscous effects such as accretion should not be
important during the simulations.

We ran two sets of simulations, one set for eccentric bound objects (planets
or stars), and one
set for stellar flybys or nearby stellar encounters.
The eccentric object simulations were run with a radial and azimuthal 
resolution of
$N_r = 150$ and $N_\theta = 450$ and 
the planet or star was begun initially at apoapse.
The flybys were run with a resolution of $N_r = 200$ and $N_\theta = 600$,
and the star was initially begun at a radius of 1.8 times $R_{max}$ 
on a parabolic orbit.
Table \ref{tab:erun} shows the characteristics of the simulations
with bound planets (simulations \#A--\#L) or stars (simulations \#M--\#P), and  
Table \ref{tab:frun} lists those involving stellar flybys (simulations \#R--\#U).
Parameters varied are the mass ratio between 
the perturber and central star, $q$ ,
the disk aspect ratio, $h/r$,
the eccentricity of the perturber, $e$, when it is bound,  
and the radius from the central star of the perturber's closest 
approach, $R_a$,  when the perturber is on a parabolic encounter.
All perturbers were assumed to be on prograde orbits.

\section{ Bound external perturbers }
We consider bound external planets and bound external stars separately.
Bound low mass objects require many orbits to move material in the disk.
The main mechanism for doing this is the excitation of spiral density waves
at Lindblad and corotation resonances (e.g., \citealt{pp3}).   
However,
an external star exerts such a strong tidal force that its Roche lobe
can reach beyond the location of most of these resonances.
Because HD 141569A has the binary HD 141569B,C 
in its vicinity \citep{weinberger2000} we can consider 
the possibility that the binary is bound to HD 141569A. 
However, HD 100546 has no known companions.  If it has a bound
companion, it is probably low mass.
Consequently,
we divide this section into two pieces, that investigating the influence
of bound low mass companions (appropriate for HD 100546)
and that that investigating the influence
of higher mass stellar companions (appropriate for HD 141569A).

\subsection{Bound external planets }

\begin{figure*}
\epsscale{1.50}
\plotone{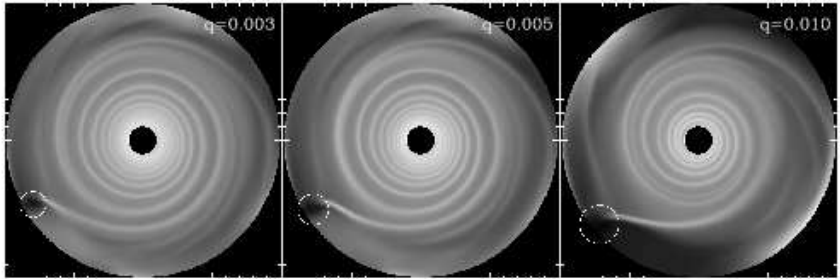}
\figcaption{
\label{fig:m_var}
The effect of planet mass on spiral structure driven by an external planet.
Simulations \#C, \#F and \#I listed in Table \ref{tab:erun} are shown.
They have planet mass ratio $q=3,5, 10\times 10^{-3}$ 
(leftmost panel to rightmost),  a disk aspect ratio $h/r=0.04$,
and planet eccentricity $e=0.25$.
The Roche radius of the planet is shown as a white circle.
The gas density is shown when the planet is at perihelion.
The planet's semi-major axis is at 1.2 times $R_{max}$ the disk outer edge.
These simulations are shown at time $t\sim 6 P_{outer}$ after the beginning 
of the simulation where $P_{outer}$ is the rotation period at the disk outer edge.
}
\end{figure*}

In Figure \ref{fig:m_var} we show the effect of different mass planets
exterior to the disk for a planet eccentricity of $e=0.25$ 
(simulations \#C, \#F, and \#I).  
We see that prominent spiral structure is driven
into these disks, similar to those exhibited in previous 
simulations of stellar binary/disk interactions
(e.g., \citealt{savonije}).  For low eccentricity, moderate mass planets,
the dominant resonance driving spiral density waves 
is the 2:1 or $m=2$ Lindblad resonance, exciting in the disk 
two strong spiral arms. 
The location of the spiral arms is
near the location of this Lindblad resonance resonance 
or at 0.76 times the semi-major
axis of the external perturber in the case of a Keplerian disk.
The dispersion relation for tightly 
wound hydrodynamic waves in a two-dimensional
gaseous disk is
$\left[m(\Omega - \Omega_p)\right]^2 = \kappa^2 + k^2 c^2$ where
$\kappa(r)$ is the epicyclic frequency, $\Omega_p$ the angular rotation
rate of the spiral pattern, $m$ is the number of arms, 
$c$ is the sound speed and $k$  the wavevector of the spiral arms.
This dispersion relation ignores the self-gravity and vertical
structure of the disk.
As expected from the dispersion relation,
the spiral structure is more open (has a higher pitch angle or lower value
of $k$) nearing the Lindblad resonance. 

While two dominant spiral arms are seen in the simulations
shown in Fig. \ref{fig:m_var}, these two arms
are not exactly symmetric.  In other words, one spiral arm
rotated about 180 degrees does not lie exactly on top of the other.
As explored by \citet{henry} in the case of M51's spiral arms, 
the interplay between different density
waves provides one possible simple explanation for such a phenomenon.
In these simulations we see a feature extending toward the location
of the planet that rotates with the planet,
and so is probably associated with the planet's co-rotation resonance.   
The asymmetries in the spiral structure are likely to be caused by the time
dependent interplay between the $m=2$ and other density waves 
(e.g., $m=1, m=3$) excited by the planet.  
Deviations from perfect bisymmetric symmetry observed in the spiral structure
of HD 100546 and HD 141569A could be explained by the interplay between 
spiral density waves.

\begin{figure*}
\epsscale{1.50}
\plotone{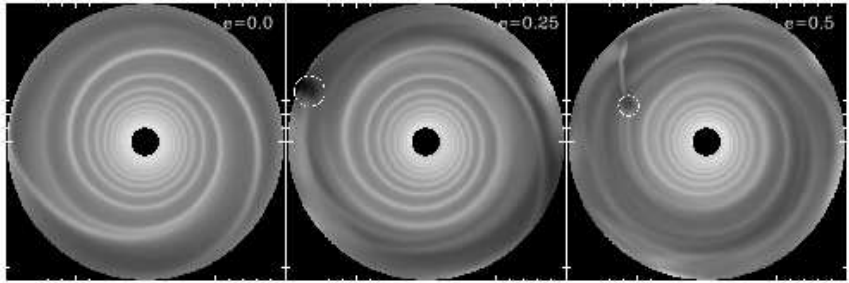}
\figcaption{
\label{fig:e_var}
The effect of planet eccentricity on spiral structure driven 
by an external planet or bound star.
The simulations \#E, \#F, and \#G are shown.  They have planet eccentricities 
$e=0.0,0.25, 0.5$ (leftmost panel to rightmost), aspect ratio $h/r=0.04$,
The simulations are shown at times with the planet near periapse.
The planet mass ratio $q=5\times 10^{-3}$.
The planet's semi-major axis is at 1.2 times $R_{max}$ the disk outer edge.
High eccentricity planets more effectively clear away gas away from the planet.
Consequently the spiral density waves driven by higher eccentricity
planets are more tightly wound.
}
\end{figure*}

\begin{figure*}
\epsscale{0.70}
\plotone{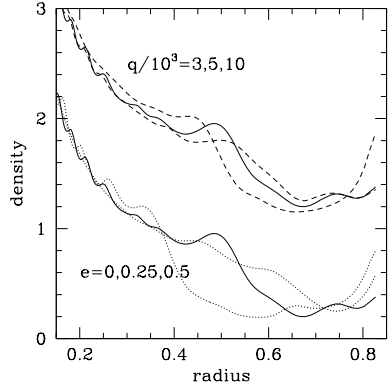}
\figcaption{
\label{fig:edge}
Edge profiles for the simulations shown in Figures \ref{fig:m_var} and \ref{fig:e_var}.
The top 3 profiles show the azimuthally averaged density as a function of radius for
the $e=0.25$, $q=3,5, 10 \times 10^{-3}$ simulations; simulations \#C, \#F and \#I.  
These are artificially offset
by a density of 1.  The bottom three density profiles refer to
those with $q=5 \times 10^{-3}$ and $e=0,0.25, 0.5$.
The two solid lines correspond to the same simulation with 
$e=0.25$ and $q=5\times 10^{-3}$.
We see that higher eccentricity planets truncate the disk at a smaller radius.
By comparing the top three profiles with the bottom three we find that
the disk edge is not as strongly dependent upon the planet mass as
on the planet eccentricity.
}
\end{figure*}

When the external planet is on an eccentric orbit, 
because additional Fourier components exist in the perturbing potential,
waves are driven at corotation resonances
as well as at Lindblad resonances \citep{GT80}.  Consequently
we would expect stronger $m=2$ spiral structure in the simulations containing
an eccentric object (e.g., \citealt{sari}).
In Figure \ref{fig:e_var} we show simulations for
different eccentricity planets (simulations \#E, \#F and \#G) but
the same semi-major axis.  
Contrary to our initial expectation, the simulations containing
higher eccentricity perturbers have lower amplitude spiral structure.
Examination of the simulations shows that
the eccentric objects more effectively truncate the disk, in part 
because the perturbing object passes deeper into the disk.
Nevertheless an $m=2$ spiral density wave is still driven into the disk, though
at lower amplitude.  When the perturbing object has eccentricity above 0.23,
the object passes within its 2:1 Lindblad resonance, 
the location where we expect 
two armed spiral density waves are excited.  Since these simulations
are shown at only $t \sim 6 P_{outer}$, we infer that 
that gas in the outer disk is quickly cleared away by the planet.
In Figure \ref{fig:edge} we show edge profiles corresponding
to the simulations shown in Figures \ref{fig:m_var} and \ref{fig:e_var}.
Except for the planet in a circular orbit, the edge of the disk 
is within the location of the 2:1 Lindblad resonance (at $r=0.76$).
We see from Fig. \ref{fig:edge} that 
the position of the disk edge is more strongly dependent on the planet
eccentricity (and the location of periapse) than on the planet mass.


One way to account for the lower amplitude spiral structure 
exhibited in the simulations containing higher eccentricity objects
is with a model where the wave is driven in low density region 
at the location of the resonance (e.g., \citealt{varniere}).
Because the waves are excited in a region of lower density, the resulting
amplitude of the waves is low even though there is more
than one resonance responsible for driving the waves.
Because the eccentric object can more effectively truncate the disk, the amplitude of
the $m=2$ waves can actually be reduced rather than increased.
In Figure \ref{fig:e_var} we also note that the spiral structure
in the simulations with higher eccentricity objects is more tightly wound.
This is expected from the dispersion relation
which implies that the waves become more tightly
wound as the distance from the resonance increases.

Figure \ref{fig:m_var},  \ref{fig:e_var}
show images at time $t \sim 6 P_{outer}$ after the beginning
of the simulation.
This time period corresponds to about
$2\times 10^4$ years and is significantly less than the age of the star.
There are two effects that reduce the density of the outer part of the disk,
viscous accretion which takes place on the viscous timescale,
and the waves themselves which carry angular momentum.
Because these simulations have disk Reynolds numbers of ${\cal R}=10^6$, 
we do not expect to see significant accretion that takes place on a viscous timescale 
($10^6$ times the orbital period) in our simulations which are only run for
100-1000 orbital periods.
However, the timescale for the spiral density waves to transfer
material away from the planet is much shorter \citep{savonije, varniere}.
From Figure \ref{fig:e_var} we can see that the 
simulations with eccentric objects
are particularly effective at driving inward gas flow in the disk edge,
whereas the simulations with zero eccentricity or low eccentricity objects 
accrete gas much more slowly.  
The disks with low eccentricity external objects 
maintain their strong $m=2$ spiral density waves to $t > 30  P_{outer}$ 
(of order $10^5$ years),
whereas those with eccentric perturbers don't.  

\begin{figure*}
\epsscale{1.50}
\plotone{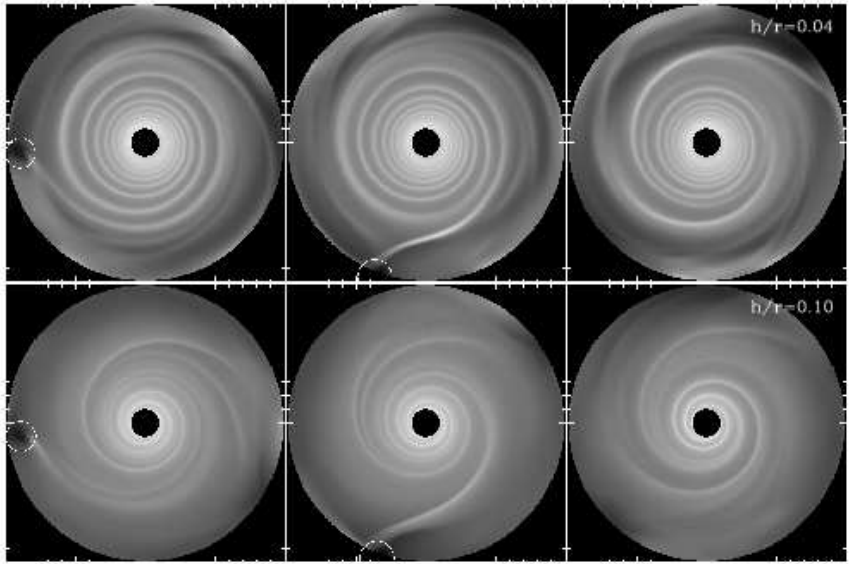}
\figcaption{
\label{fig:h_var}
The effect of the sound speed on spiral structure driven by an 
external planet or bound star.
Two simulations, \#F and \#L, 
are shown for times when the planet is at periapse (left
panels), at apoapse (right hand panels) 
and at an intermediate time (central panels).
The planets have eccentricities $e=0.25$, 
and planet mass ratio $q=5\times 10^{-3}$.
The simulations shown in top panels have a disk aspect ratio of $h/r=0.04$ and
those shown in bottom panels $h/r=0.10$.  
Thicker (or higher sound speed) 
disks exhibit lower amplitude and more open spiral structure.
}
\end{figure*}

From the dispersion relation of tightly wound arms, we expect that thicker
disks with a higher sound speed would exhibit more open spiral structure 
(as previously seen in the simulations of \citealt{savonije}).
Figure \ref{fig:h_var}  shows how the disk aspect ratio 
(which sets  the sound speed) affects the morphology.  
This figure shows the same simulations at different orbital phases;
the leftmost panels show the perturber at periapse and the right
most panels show the perturber at apoapse.
We see from this figure that the 
spiral density waves excited do tend to be more tightly wound 
when the sound speed is lower and the disk thinner.   
We also find that in the larger scale height disks, the slope of
the disk edge is shallower, suggesting that the resonant excitation 
of the spiral density waves
arises from a larger region in the disk \citep{artymowicz94, pp3}.

From Figure \ref{fig:m_var} we see that low mass (Jupiter sized), low eccentricity 
bound perturbers can excite two armed moderate amplitude
spiral arms.  This morphology is similar
to that exhibited by the disk of HD 100546.
Because of the short timescale (a few rotation periods at
the outer edge of the disk) that gas is cleared to within
the 2:1 Lindblad resonance,
strong observed open 2-armed spiral structure as seen in the disk of HD 100546 
is unlikely to be excited by a highly eccentric ($e \gtrsim 0.20$) 
bound planetary perturber. 
On a timescale of a few rotation periods, the edge of the disk
moves sufficiently away from the perturber that strong open two-armed 
structure is no longer evident.

We now compare the morphology of the simulations to that seen HD 100546.
The disk of HD 100546 exhibits 2 arms that are
more open at larger radii than at smaller radii.  
The simulated disks shown in Fig.\ref{fig:m_var} also show 
spiral structure that is more open at larger radii and more tightly
wound at smaller radii.  
The simulations also show more tightly wound structure at smaller radius
that is not detected in the image of HD 100546. This could be in part
because of the increased level of light from the central
star which would make it difficult to detect fine structure in the disk.
Alternatively, the disk of HD 100546 may be comprised of planetesimals;
a particle disk would damp tightly wound waves.
For objects on circular orbits, 
we find that the amplitude of the waves drops below about 15\%,  
(as observed for HD 100546),
when the mass ratio $q \lesssim 0.005$.  If a bound perturber is responsible
for the spiral structure in the disk of HD 100546, our simulations
suggest that the perturber must have
mass above 10 Jupiter masses.  If the disk is thicker a higher
mass perturber would be required.
We will use these limits in Section 4 when
we discuss observational constraints on these scenarios.

\subsection{Bound external stars }
For HD 141569A, the likely perturber is the companion binary HD 141569B,C,
consisting of an M2 and an M4 star,
located  $9''$ away from HD 141569A \citep{weinberger2000, augereau2004}.
We estimate a mass ratio for the binary compared to HD 141569A of 
$q\sim 0.2$ based on the low mass stellar models 
by \citet{baraffe2002} and effective
temperature scale by \citep{luhman}, consistent with $q =0.2$ adopted
by \citet{augereau2004}.
The amplitude of the outer spiral structure (at radii
greater than 300 AU) in the disk of HD 141569A is much higher
than that of HD 100546 and the nearby binary provides a likely candidate
responsible for driving it.  However $q=0.2$ is well above the mass
ratios considered for the previous simulations.   To explore the
way that spirals structure could be excited in the disk of 
HD 141569A, we ran a set
of simulations with higher mass external bound perturber with $q=0.2$,
(see Table \ref{tab:erun}).  

\begin{figure*}
\epsscale{1.50}
\plotone{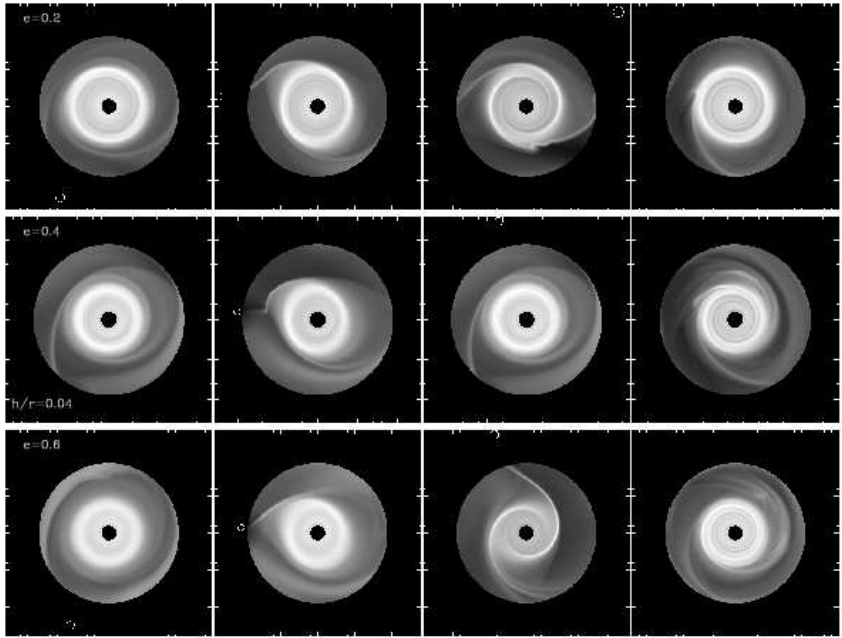}
\figcaption{
\label{fig:hd_evar}
The role of eccentricity on the disk morphology for
external bound stellar mass companions.
These simulations (\# M, \#N and \#O) 
have mass ratio $q=0.2$, 
and disk aspect ratio $h/r = 0.04$.
The position of periapse is approximately 
1.1 times the outer disk radius in all simulations.
The top panels have perturber eccentricity $e=0.2$, the middle
ones $e=0.4$, and the bottom ones $e=0.6$.
Each panel (left to right) shows simulations just before periapse 
(leftmost panel), at periapse, after periapse and nearly at apoapse (rightmost 
panel).  The small circles show the location of the perturber.
Just following periapse, asymmetric spiral arms are pulled out of the disk.
The perturber truncates and compresses the disk, enforcing a higher gas density 
near the edge and a sharper edge profile.  
Spiral density waves driven into the disk cause elevated
accretion episodes in the disk interior.
}
\end{figure*}

In Figure \ref{fig:hd_evar} we show simulations for different binary
eccentricities (\#M, \#N and \#O).  
Simulations \#M, \#N are shown approximately 5.5 binary orbital rotation periods  
(BC about A) after the start of the simulation and simulation \#O is
shown approximately 4.5 orbital rotation periods 
after the start of the simulation.   
It is computationaly expensive for us to run finely gridded 
hydro simulations for significantly longer time periods.
The disk is truncated by the binary on its first few passages, 
but each subsequent close encounter excites spiral
structure.    The size of the disk is consistent with truncation
at the Roche radius of the binary when the binary is at periapse. 
Consequently the disk outer edge is well within the location of the binary's 
2:1 Lindblad resonance.
We see in the simulations that following the first few passages
the disk outer edge is much higher density than the initial disk density.
The binary has not only truncated the disk, and induced a sharp outer edge
but has also compressed it.  The high edge density,
could be seen as a ring of material, and the lack of gas interior to the ring
(e.g., \citealt{clampin})
might be mistakenly interpreted as evidence that material has been cleared
out by another process such as an interior planet.

Observations of HD141569A suggest that the disk is cleared between 150 and
200 AU \citep{mouillet}.  Clearing at these inner radii is probably not
caused by the perturbations from the binary HD141569B,C, and so is probably
induced by internal processes, such as planets (see discussion by 
\citealt{mouillet}).
The inner boundary of our simulation is sufficiently distant from
the truncated edge that it should not have affected
the outer disk structure.   Spiral waves driven
in these simulations are damped before they reach the inner regions
of the simulated disk.  So the presence of the clearing 
at 200 AU should not be able to strongly affect the morphology observed
in the outer disk at 300--450 AU. 

In Figure \ref{fig:hd_evar} we show panels preceding and following
the binary's periapse.  Before periapse we see the disk is nearly
axisymmetric and is quiescent. It lacks strong spiral structure.  
However at periapse the tidal force from the binary begins to
pull two arms from the disk.  Since one arm is directly
excited by the binary and the other excited by the recoil of HD 141569A 
\citep{pfalzner},
there is an asymmetry between the spiral arms.
The disk is lopsided when the binary is near periapse.
The asymmetric spiral structure,
the lopsided disk, and the spiral arm pointing toward
the perturber are features that are exhibited by the disk
of HD 141569A as seen in scattered light \citep{augereau99b,clampin}.
We find that outer
spiral arms are less strongly excited by a binary on a more highly
eccentric orbit.  This is likely because the binary on a 
higher eccentricity orbit is moving somewhat faster at periapse
than one at a lower eccentricity,  resulting in a weaker impulse. 

\begin{figure*}
\epsscale{1.50}
\plotone{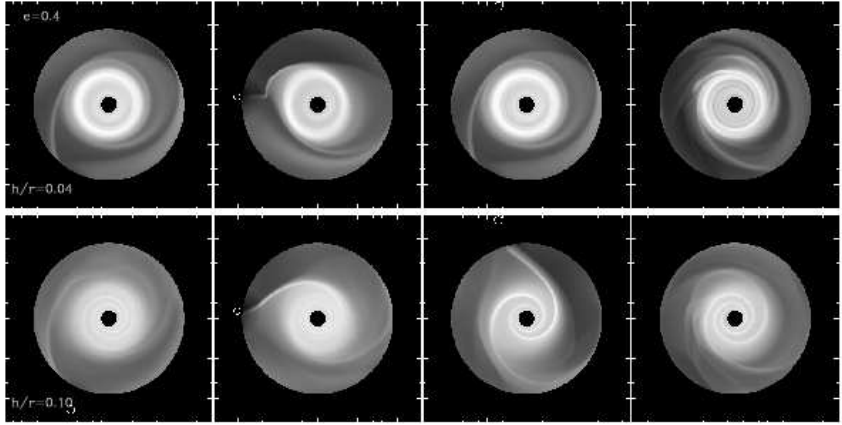}
\figcaption{
\label{fig:hd_hvar}
The role of disk thickness for
external bound stellar mass companions.
These simulations have mass ratio $q=0.2$,  and
perturber eccentricity $e=0.4$.
The disk aspect ratio is $h/r = 0.04$ for the top panels (simulation \#N) and
$h/r = 0.1$ for the bottom panels (simulation \#P).
Each panel (left to right) shows simulations just before periapse 
(leftmost panels), at periapse, after periapse and nearly at apoapse.
The disk edge profile is smoother when the disk is thicker
and the spiral structure is more open.
}
\end{figure*}

We investigate the role of disk thickness by displaying 
in Figure \ref{fig:hd_hvar}
two simulations (\#N and \#P) that differ only in the disk aspect ratio.
The outer spiral arms excited by the encounter with the binary
are more open in the thicker disk, however the amplitude
of the features is lower, particularly in the inner region.
Also the outer arms point directly toward the binary, rather than leading
the binary as is seen in HD 141569A.

\begin{figure*}
\epsscale{1.50}
\plotone{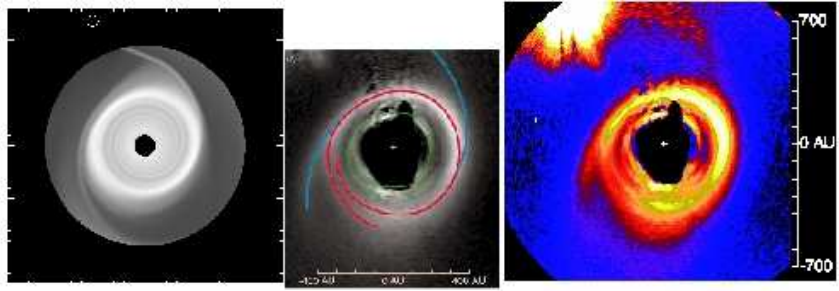}
\figcaption{
\label{fig:color}
Comparison between a simulation with $q=0.2$, $h/r=0.04$, $e=0.2$, 
simulation \#R, 
with the binary companion HD141569B,C near periapse on a prograde orbit, 
and the observed morphology of HD 141569A. 
The images in the middle and on the right are from \citep{clampin}.
Our simulation does a reasonable job of accounting for the outer spiral
structure, the location of the disk edge, matches reasonably well
the position of the binary (denoted with a white circle), 
and predicts an asymmetry in the location of the 
outermost two arms that is seen in the disk of HD141569A.
Much of the structure of the outer disk may have been
excited by the tidal force of the binary HD141569B,C. 
However our simulation fails to exactly match the shape of the outermost 
spiral features, and the lopsided nature of the disk.  
We suspect that a better 3D simulation would be required to do this.
The images of HD141569A are shown with north to the left and 
west on top as by \citet{clampin} and \citet{augereau2004}.
}
\end{figure*}

We found the closest correspondence between the observed morphology
and that seen in the simulations with binary eccentricity
$e\sim 0.2$,
a thin disk ($h/r \sim 0.04$) and binary approaching periapse (simulation \#R).
This simulation is shown in figure \ref{fig:color} along with
the observed morphology pointed out by \citep{clampin}.
We support the scenario proposed by \citet{augereau2004}  in which
the disk structure is influenced by perturbations from the binary.
Our simulation is successful at producing open and high amplitude two armed spiral
structure at the outer edge of a sharp disk, similar
to that observed in the HST images at radii greater than 300 AU.  
It successfully accounts for the disk truncation. The simulation also exhibits
some assymetries that are similar to those observed.  
The simulated northern spiral arm is not 180$^\circ$ degrees away from
the southern spiral arm, but shifted or rotated westward so it is closer
to the binary than it would be if it were an 
image of the southern spiral arm rotated by 180$^\circ$. 
The simulated disk edge is lopsided, with the northern edge thicker and 
more distant from the star than the southern edge.  
However, the observed disk edge is brighter on the 
southwestern side than
on the northeastern side and this brightness difference 
is not seen in our
simulation.  The northern spiral arm is further east and
the southern arm is further west 
in the simulations than the observed ones.

While we have attempted to match the outermost structure of the disk,
\citet{augereau2004} concentrated primarily on the structure of the disk edge.
Our simulation accounts for the outer spiral structure but is less good
at matching the assymetry or lopsided appearance of the bright rim of 
the disk.  This assymetry was well modelled with the secular model
of \citet{augereau2004} though their simulations did not exhibit the outer
spiral arms that are evident in our simulations. 
\citet{augereau2004} and our simulation also differ in the predicted
location of the binary periapse.  
Our simulation has binary periapse near the current location
of the binary HD141569B,C, whereas \citet{augereau2004} suggest that
the periapse is at position angle $\sim -130^\circ$ or on 
the southwestern side of the disk.

To account for the outer spiral arms we have assumed 
that the binary is in a prograde orbit. 
A prograde encounter
allows open high amplitude spiral arms to be drawn out of the disk 
when the binary is near periapse.
To truncate the disk, a much closer periapse would
be required with a retrograde orbit for the binary.  Retrograde encounters
are also much less effective at pulling out spiral arms near closest approach.
The model of \citet{augereau2004} did not specify whether the binary
was on a prograde or retrograde orbit.  Periapse is on the southwestern 
side of the disk in their model, 
so the binary is either post-periapse and retrograde
or pre-periapse and on a prograde orbit. 
Since the phenomena they illustrate and describe in the disk
was caused by secular perturbations, 
the direction of the binary orbit may not be important in their model.
Our simulations suggest that much of the outer structure in
the disk is transient and excited specifically by the current close encounter.
We find that two armed open spiral arms are best excited 
just past periapse for a prograde binary orbit. 
This presents a problem for the model of \citet{augereau2004} since
neither possibility (post-periapse and retrograde or pre-periapse
and prograde) would be capable of
exciting the outer spiral structure.  
Modifications of the secular model of \citet{augereau2004}
will be required to account for the outer spiral structure. 
We note that the outer spiral arm connecting
the HD141569A disk to the stellar companions was only
barely detected in the images presented by \citet{clampin}, whereas
that on the north was detected by both \citet{clampin} and 
\citet{mouillet}.  
If future observations fail to confirm the presence of the southern
arm then a different scenario or very different orbital parameters
would be required to account for only the nothern arm.

Our simulation does exhibit some lopsidedness in the disk,
even though the binary eccentricity we adopted was not as high
as that adopted by \citet{augereau2004}.
The simulations of \citet{augereau2004} found that 
a highly eccentric binary was required
to account for the disk asymmetry \citep{augereau2004} on long timescales.  
We found that a binary on a more eccentric orbit excited slightly
more assymetry in the disk near periapse.
However, the disk tended to be offset toward the
perturber rather than toward the west, as observed \citet{mouillet}.
In our simulations,
the assymetry damps as the binary recedes following periapse.
The hydro simulations explored here are unlikely to maintain the elliptical orbits
exhibited by the particle simulations of \citep{augereau2004}.
It's possible that a particle simulation would provide a better
match to the physical state of the disk on longer timescalse than
our hydro simulation.
Our simulations exhibit weak spiral waves driven inside 200 AU that 
are not observed in the observed images of HD 141569A.  
Because they are tightly wound, they are similar to sound waves,
and would not be present if the disk is comprised of planetesimals.

While our simulation provides a reasonable match to the observed
outer morphology we found previously that
lower eccentricity binaries and colder disks exhibit higher amplitude
and finer structure in the disk. 
This degeneracy makes it difficult for us
to finely discriminate between models and so estimate the eccentricity
of the binary's orbit.  While we find somewhat better correspondence between
the simulations and observed outer spiral arm morphology with a lower eccentricity
binary and relatively thin disk, we cannot exclude the possibility
that the binary could have a higher eccentricity, as suggested
by \citet{augereau2004}.

The chosen simulation, \#R,  shown Figure \ref{fig:color}, 
predicts the outermost 2 spiral arms,
matches reasonably well
the position of the binary, accounts for the steep disk edge profile,
and predicts an asymmetry
in the outermost two arms that is seen.  However we fail
to exactly match the shape of the outermost arms and the structure
in the disk edge.
Our previous figures showed that this issue cannot be resolved
with a binary with higher eccentricity or with a thicker disk
because these simulations tend to have less well defined inner disk
structure, though 
because of the degeneracy between binary eccentricity and disk thickness
we cannot tightly constrain both quantities.
We note that our simulations were carried out in 2D, and it is 
likely that the binary is not exactly in the same
plane as HD 141569A's disk.
The worst discrepancies between our simulations
and the observed disk would be in the outer regions where the
tidal force from the binary on the disk is largest.
By carrying out 3D simulations, it may be possible to more exactly match
the morphology and so more tightly constrain both the thickness
of the disk and eccentricity of the binary orbit from the
disk's morphology.  We have assumed here that a hydrodynamic simulation
provides a good approximation to the disk of HD 141569A.  
However, if this disk is a debris disk, particle simulations that 
include collisions might provide a better physical model.

Assuming that the binary HD141569B,C is at periapse and an
eccentricity of 0.2, the period of the orbit is a few thousand years.
Based on an age estimate of 5Myrs for this system,
we estimate that
a few thousand orbits could have occurred since the birth
of the system.  \citet{augereau2004} inferred a high orbit eccentricity
based on the timescale for the maintainance of a lopsided disk.
If the orbit of the binary is stable, then many periapse encounters could
have taken place, particularly if the binary is not on a highly
eccentric orbit.  While a few periapse encounters 
are required to truncate the
disk, we expect that repeated encounters could more strongly truncate
the disk.  As the disk decreases in size, the binary
would pull out fainter and fainter arms
from the edge of HD141569A's disk.   Nevertheless we have proposed here
that the outer spiral arms observed in the disk by \citet{clampin}
are a result of tidal excitation from the binary HD141569B,C.
We note that the wide separation of HD141569B,C from HD141569A implies that
this system may eventually be disrupted altogether by the passage
of field stars.  The probability of such a field star perturbing
the HD141569A,B,C system can be
estimated as $P \sim n v A \Delta t$ where $n$ is the number density
of stars ($\sim 10 {\rm pc}^{-3}$), $v$ is the field star velocity dispersion
or $\sim 10$km/s,  $A$ is the cross section of the binary orbit 
($\sim  \pi (1000 AU)^2$) and $\Delta t$ is the timescale (similar to
the age of the system).
Based on these quantities, we estimate
that the HD141569A,B,C system has a probability of 
$\sim 5\%$ of a field star encounter
during the lifetime of the system.
If the binary is on a more highly eccentric orbit then the probability
would be higher.  One interesting possiblity is that the disk
of HD141569A has only recently been truncated by the binary orbit.
If the binary is on an eccentric orbit, 
its orbit could have been perturbed
by a nearby field or birth cluster star.

\subsection{Excitation of spiral structure by flybys}

In this section we consider stellar encounters as possible
causes for spiral structure in circumstellar disks.  This is 
relevant for HD 100546 which does not have any known nearby companions 
\citep{grady}.

\begin{figure*}
\epsscale{1.80}
\plotone{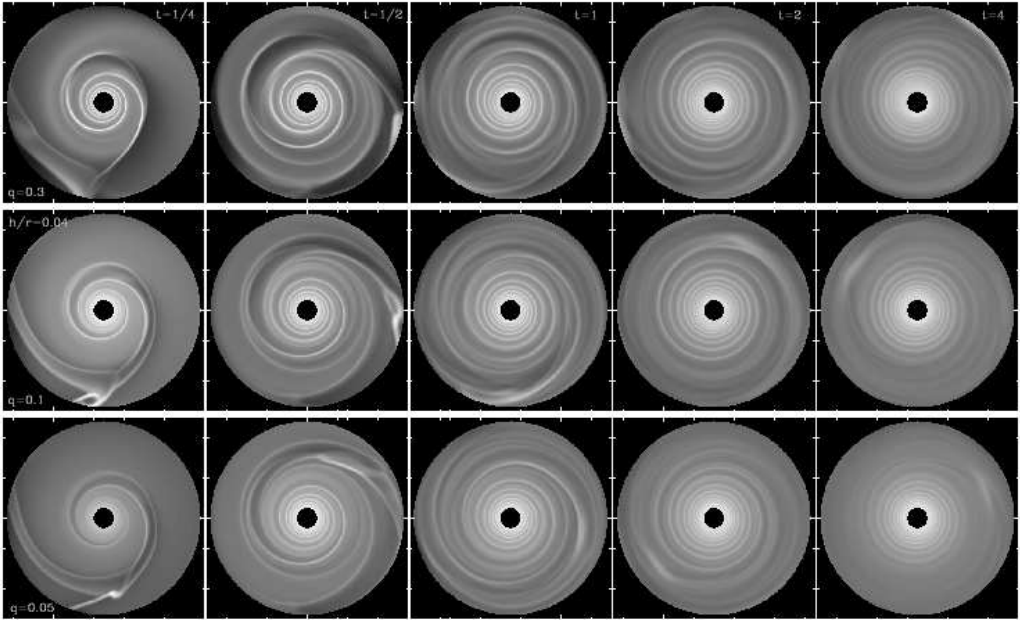}
\figcaption{
\label{fig:F_mvar}
The effect of stellar mass on spiral structure excited by
a recent stellar encounter.
Simulations are shown at $t=1/4,1/2,1,2$ and 4 times $P_{outer}$
(the rotation period at the disk outer edge)  from left to right
after a star of mass ratio $q=0.3, 0.1$, and $0.05$
undergoes its closest approach to the central star.
Each row corresponds to a simulation with different mass ratio (simulations
\#R, \#S and \#T respectively).
The position of closest approach is half
the radius of the disk outer edge and the star 
is assumed to be on a parabolic trajectory.
For these simulations the disk Reynolds number ${\cal R} = 10^6$, 
and the disk aspect ratio $h/r=0.04$.  
If the perturber's mass is below $\sim 0.1$ times that
of the star, only one spiral arm is dominant.
The short timescale of the excited spiral structure places a limit
on the proximity of a perturber that  could
have been responsible for driving the spiral
structure in the disk of HD 100546.
As a guideline, the edge of the disk should be compared to the size
of HD 100546's disk or $\sim 300$AU.
}
\end{figure*}

In Figure \ref{fig:F_mvar} we show simulations for parabolic stellar
flybys as a function of
time after the closest approach.  Fig \ref{fig:F_mvar} shows simulations for
stellar mass ratios of $q=0.3, 0.1$ and $0.05$, simulations \#R, \#S and \#T respectively.   
We ran these simulations
so that the star passed well within the disk to minimize  
the effect of the outer boundary.   
Nevertheless at late times, the disk is clearly affected by shocks associated
with the outer boundary.  Our code is complementary to previous
studies because it is good at showing spiral structure,
but not as accurate as N-body or SPH codes at modeling elongated tidal arms 
(as simulated by \citealt{pfalzner,boffin}).
Initially the gas disk in our simulations exhibits a global two-armed spiral.
The size of the secondary arm appears initially to depend on the mass ratio,
though at later times the morphology is less assymetric or lop-sided. 
\citet{pfalzner} showed that the secondary arm
is dependent upon the recoil of the central star and our simulations
show the same dependence.  Since HD 100546 exhibits
two-arms, the perturber (if a flyby) probably did not have 
a small mass ratio.  From our simulations we estimate
$q \gtrsim 0.1$, corresponding to $M \gtrsim 0.2 M_\odot$, 
since lower mass perturbers fail to excite two strong spiral arms.

After 4 rotation
periods (at the outer disk edge) little strong spiral structure remains.
This short timescale is consistent with the results of previous simulations 
\citep{pfalzner},
and allows us to place limits on the time since the closest approach.
The spiral structure HD 100546 is likely to have been excited less than
one rotation period (at the outer disk edge) ago, which is less than 3000 years
ago based on the rotation period at the edge of HD 100546's disk.
For an object traveling at 10 km/s this would correspond to 
a distance on the sky of 1 arcminute.

\begin{figure*}
\epsscale{1.80}
\plotone{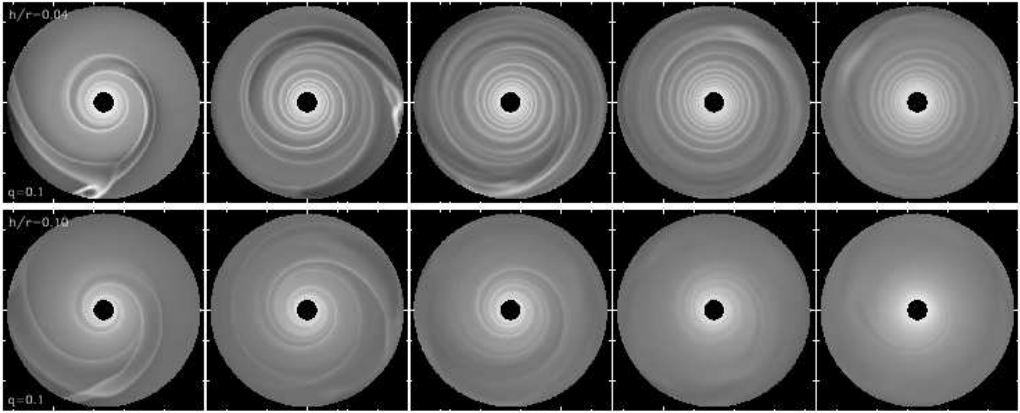}
\figcaption{
\label{fig:F_hvar}
The effect of disk thickness on the morphology of spiral structure
excited by a recent stellar encounter.
The top panels show a simulation with a disk aspect ratio $h/r =0.04$, 
simulation \#S,  compared
to that with  $h/r =0.10$, simulation \#U,  which is shown on the bottom panels.  
Each panel (left to right) corresponds to
a different time after the closest approach (with times the same
as those in Figure \ref{fig:F_mvar}).
For these simulations the mass ratio $q=0.1$.
Thicker disks exhibit lower 
amplitude and less long lived prominent spiral structure
following a flyby.
}
\end{figure*}

Since the pitch angle and wave travel time depends on the disk sound speed, 
we ran a comparison simulation with $h/r = 0.1$ (simulation \#U) that is shown in 
Figure \ref{fig:F_hvar}. 
We see that the amplitude of the spiral density waves is 
reduced in the thicker disk, even though they are more open.
Because the amplitude of the spiral structure is lower, a higher mass perturber
that encountered HD 100546 more recently would be required to 
account for the observed spiral structure. 

\begin{figure*}
\epsscale{1.50}
\plotone{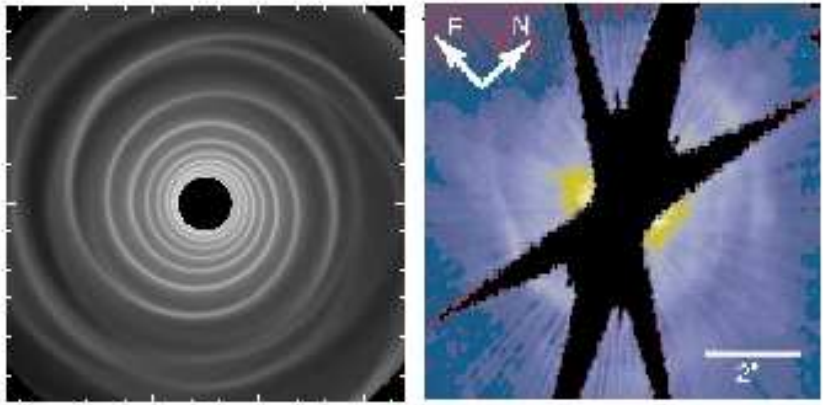}
\figcaption{
\label{fig:color2}
Comparison between a flyby simulation with $q=0.3$, $h/r=0.04$, simulation \#R, 
and the observed morphology of HD 100546.  The simulation is shown
about 3000 years after the encounter.
The image on the right is the STIS image by \citep{grady}.
As long as the perturber is $\gtrsim 0.1$ times the mass of
the star two arms will be excited.  The pitch angle of the arms
is related to the time since the perturber reached periapse.
We have found that many simulations are capable of exhibiting open
two armed structure similar to that seen in the disk of HD 100546.  
To constrain
the nature of the excitation, it is necessary to identify the object
responsible for the excitation.  A flyby excites short-lived spiral structure 
and need not sharply truncate the disk. 
}
\end{figure*}

In Figure \ref{fig:color2} we show a comparison between
an encounter and the observed morphology of HD 100546. The simulation
is shown about 3000 years ($t =1 P_{outer}$) after the encounter
for a mass ratio $q=0.3$ and with disk aspect ratio $h/r=0.04$ (simulation \#R).
The simulation exhibits an asymmetry in the spiral structure similar
to that observed, though the easternmost spiral arm is not as open
as the observed one.  This figure demonstrates that 
it is easy  (and so not particularly meaningful)
to simulate spiral structure similar to that observed in the disk of
HD 100546. 
We must identify the actual object responsible for exciting the 
spiral structure to constrain this scenario further.



\section{Observational limits on HD 100546's perturbing object}

As discussed by \citep{grady,augereau01} 
the STIS, NICMOS and ADONIS observations
strongly constrain the types of nearby objects that could be associated 
with HD 100546.  The nearby point sources detected in the STIS images
are all too blue to be associated with the HD~100546 system, and so are likely
to be background stellar sources \citep{grady}.   This puts strong limits
on both of our scenarios.  If a flyby caused the spiral structure,
then the object responsible must be outside the field of view of the STIS
image ($\gtrsim 20''$ away from HD 100546).
Alternatively if a low mass bound object causes the spiral structure,
then the mass required to excite the spiral structure must be low enough
so that it would not have been detected in the existing images.
Our simulations suggest that a bound mass 
at least as large as 0.01 times the mass of
the star is needed to match the amplitude of the spiral structure. This would
correspond to a mass of $0.02M_\odot$ or $20 M_J$ (see Section 3.1). 

The division between bound perturber and flyby is important.
A bound perturber will make more than one encounter over a timescale
of a few thousand years.  Since a
first encounter disrupts the disk, the remaining disk may not 
exhibit strong spiral structure during subsequent approaches.
Since a flyby disrupts a previously
quiescent disk, strong outer spiral structure can be excited, but it would
be transient, only lasting a few thousand years. 
Afterward the dirupted disk could extend well past the pericenter
of the encounter.

The lack of detected associated objects (point sources)
in the NICMOS and STIS images 
can be used to estimate an upper limit on the mass of possible objects
near HD 100546.
The models of \cite{baraffe2002} predict that 5$M_J$ and 10$M_J$ mass objects 
that are 10 Myr old
have absolute H magnitudes of 15.3 and 11.8 respectively.
These correspond to H magnitudes of 20.3 and 16.8 for objects
at the distance of HD 100546.
The NICMOS image of \citep{augereau01} detected a star of H magnitude 16.1
at a distance of 4'' from HD 100546, suggesting that objects
somewhat fainter could have been detected nearer HD 100546.
The disk brightness drops to 17.5 mag/arcsecond$^2$ in H band at a radius
of 3'' from the star \citep{augereau01}.  Considering the NICMOS point spread
function, a star of magnitude 19-20 could have been detected against the disk
at a radius of 3'' from the star.
If we adopt $10 M_J$ ($0.01 M_\odot$)
 as a limiting  mass for a coeval object that could
have been detected in the STIS and NICMOS imaging to a radius of 3'', 
then we can limit the possible bound companions of HD100546 within
$\sim 300 AU$ to a mass ratio $q \lesssim 0.005$.
The perturber mass limit suggested by the simulations, $M> 0.02 M_\odot$ is 
in contradiction to the limits placed on possible companions by
the lack of associated objects detected in HST/STIS and NICMOS observations.
We conclude that a bound low mass companion is not a good explanation for
the observed spiral structure in the disk of HD 100546.


We now consider observational constraints on objects
that could have previously encountered HD 100546.
USNO-B1.0 is an all-sky catalog complete to a visual magnitude of 21
that presents positions accurate to 0.2'' and proper motions measured across 
50 years of photographic sky surveys \citep{usnob}.
In the vicinity of HD 100546 (within 2 arcminutes)
the USNO-B catalog (using surveys during 1976, 1984 and 1991) 
yields one nearby object with a similar proper motion to HD 100546,
USNO-B1.0 ID \#0197-0270043, that is 54" arcseconds away from  HD 100546.
This star with $M_R=17.4$ does not coincide with any 2MASS object, and it is quite blue
($B-R\sim 0$)
suggesting that it is a white dwarf rather than a low mass main sequence star.
Consequently it is unlikely to be the same age as HD 100546 and so is unlikely
to be associated with the star.  
Inspection of the USNO-B digitized images suggests that the proper motion
is uncertain due to two nearby stars,  one which
does have a 2MASS counterpart.

We also searched the USNO-B1.0 catalog for objects that might not be
associated with HD 100546 but could
have intersected the path of this star on the sky.
We find no moving candidates that are red enough to
be at the same distance as HD 100546 from us and that could
have approached (within 15,000 AU)  the path of HD 100546 
in the past 9000 years.
Both the 2MASS and USNO-B catalog contain numerous red faint sources within
a few arcminutes of HD 100546 that could either  be candidates for
low mass stellar bound companions or low mass stars that  
could have encountered the star in the past
few thousand years.  We failed to find an extremely red object that 
was seen in 2MASS (limiting magnitude $K\sim 15$) 
and was not seen in the USNO-B surveys (limiting magnitude $V \sim 21$).
The 2MASS all sky survey point source catalog contains 23 stars with
near-infrared color
$J-K > 0.9$ within 2' of HD 100546.  These are candidate low mass
main sequence stars at the distance of HD 100546 lacking measured proper motions
from the USNO-B catalog. 
Our failure to find a catalog candidate for the object responsible suggests
that it must be a low mass object.  Stars later than M6 at the distance of HD~100546 
(100 pc)
would not be visible in the USNO-B surveys.  
Brown dwarfs at the distance of HD 100546 would not be seen in the 2MASS survey.

The low proper motion of HD 100546  with
$\mu$RA  $=-38$mas/yr and 
$\mu$DEC $= -2$mas/yr makes it difficult to pinpoint 
faint stars at the same proper motion using ground based data across 
only a 15 year baseline (USNO-B1.0
is based on surveys in 1976, 1984 and 1991 in this region), 
during which time HD 100546 would have moved by only $0.6"$.

Stellar collisions or flybys in the field are unlikely  
(e.g. \citealt{kalas01}).  Assuming a stellar density of 10 stars pc$^{-3}$ 
and a stellar velocity dispersion of 10km~s$^{-1}$ 
typical of the Milky Way density in midplane,
and a cross section of $\pi (600 {\rm AU})^2$ for HD 100546's disk,
we would estimate a probability of $0.03$ that a low mass star
could have approached within 600 AU of HD 100546
in the past 10 Myrs.    The probability is only $3\times 10^{-6}$ that
a field star approached HD 100546 in the past 10,000 years, the upper
limit on the time since the encounter we have inferred based on the
the morphology of the spiral structure.
However, the birth cluster of
HD 100546 may have provided a denser stellar environment, increasing
the probability of a stellar collision \citep{kalas01,adams}.
As pointed out by \citet{vieira}, HD 100546 is located just
to the south east of the dark filament DC296.2-7.9.
Through their study of the extinction in this cloud \citet{vieira,corradi} find 
this cloud is likely to be just past HD 100546,
110--130 pc away, and so could be associated with HD 100546.
The statistics of star colors in the region 
suggest a moderate mininum extinction of $A_V \sim 0.5$ is provided
by the cloud \citep{vieira}.
From inspection of 2MASS images near HD 100546 we find  
no obvious stellar cluster associated
with the dark filament.  Unless a region of higher stellar density is found
within a few arcminutes of HD 100546, the encounter would have
been an exceedingly improbable and exceptional event.


\section{Summary and Discussion}

In this paper we have used 2D hydrodynamical simulations of disks perturbed 
externally by stars or planets to  investigate possible scenarios
which can account  for outer spiral structure observed
in the circumstellar dusty disks of HD 100546 and HD 141569A.  We consider two
types of external perturbers, those in orbit about
the central star containing the disk and 
those involving a recent stellar encounter or flyby.  Both scenarios
produce morphology similar to that observed, moderately open 
2-armed spiral structure.   

Our simulations exhibit some trends. 
Thicker disks tend to have lower amplitude and more openly wound
spiral density waves.
Eccentric planetary perturbers more efficiently clear gas away from 
the perturber.
At earlier times in the simulations, eccentric planetary perturbers 
excite strong and open spiral structure, however after a few
rotation periods,
the edge of the disk moves inward, and the waves excited become
more tightly wound and amplitude drops.
The two armed structure excited tends to be asymmetric,  a phenomena
we attribute to the interplay between $m=2$ 
and additional spiral density waves at different angular rotation rates
that are also excited by the perturber.

When the bound object is higher (stellar) mass, the disk is truncated
at the Roche lobe of the companion.
Each time the companion approaches periapse,
tidal arms can be pulled out from the disk and spiral density
waves driven into the disk.   
We see from the simulations that the disk is not only truncated
by the binary 
but also can be compressed, causing a high gas density at the disk edge resulting
in the appearance of a gas ring.

We find that the outermost two spiral
arms of HD 141569A's disk are qualitatively reproduced 
by a perturbation
from its nearby binary companion HD 141569B,C, confirming the scenario
proposed by \citet{augereau2004}.  
However, we find closer morphological correspondence with thinner disks 
($h/r \sim 0.04$), moderately $e \sim 0.2$, 
but not highly eccentric orbits for the external binary with the binary
located near the periapse of its orbit.
Because a higher binary eccentricity causes a similar effect
on the morphology as a thicker disk, these two parameters
are redundant and we cannot exclude the possibility that the binary
eccentricity is higher.
Our simulation accounts for only some of the asymmetries
in the inner region (200-300 AU). It matches reasonably well
the position of the binary, and predicts an asymmetry
in the outermost two arms that is seen in the disk of HD 141569A.
However, our simulations also are not as lopsided as the observed disk
and do not exhibit the high density on the southwestern side
of the disk that  was seen in the observations by \citet{clampin}.
Our simulations do not do as good a job as the secular
collisionless particle simulations of \citet{augereau2004}
at accounting for the lopsided shape of the disk edge, though
those simulations do not exhibit outer spiral arms.
Our simulations also fail to exactly match the shape of the outermost
spiral features.  We suspect that a 3D simulation would be required
to do this.  Our simulations also exhibit an enhanced density
at the disk edge (similar to that observed), 
and episodes of high accretion in the inner
disk following the binary's close passage.    In a planetesimal
dominated disk (rather than a hydro one), 
these episodes could correspond to episodes of increased collisions 
and dust production \citep{kenyon}.

The current uncertainty in the composition of these disks hampers
our ability to simulate them.    These disks are suspected to 
contain planetesimals, but collisions are also expected to
be important over the lifetime of the disks.  Here, we have opted
to carry out 2D hydrodynamic simulations as a first step.  
In future 3D particle
simulation which include collisions would provide a better physical match
to the suspected properties of debris disks.  
The simulations of \citet{augereau2004} investigate long
timescale secular perturbations and suggest that much 
of the morphology of this disk has been caused by repeated
encounters with the binary HD 141569B,C.  In contrast
our simulations suggest that some of the structure (the outer
spiral arms) in HD141569A's
disk has been caused by a recent encounter with the binary.  
From attempting to match the orientation of the
the tidally excited spiral arms, we infer that the binary HD141569B,C is
near periapse and that the binary is on a prograde orbit.  
Our estimated location
periapse differs from that predicted by \citet{augereau2004}.
Future numerical work could attempt to account for both
the outer spiral arms (as done here) as well as 
the disk edge structure (as better done by \citealt{augereau2004}).

In the case of HD 100546, a low mass planetary perturber in a circular orbit
might cause low amplitude two-armed spiral structure that 
is more open at larger radii than smaller radii. 
However, to match the observed spiral amplitude, $\sim 15\%$ , 
the bound object must have a mass greater than 
$M > 0.02 M_\odot$, in contradiction to the limits placed
by the lack of detection of possible associated objects 
in the NICMOS and STIS images of \citet{grady,augereau01},
($M \lesssim 0.01 M_\odot$ if the object is the same age as
HD 100546).  A recent stellar encounter 
might also be responsible for the spiral structure.
Our simulations suggest that an star of mass about $0.1 M_\odot$
passing within 600 AU of HD 100546 within a few thousand
years ago could account for the spiral structure. 
The timescale since the encounter is long enough that the star
could be outside the field of view of the STIS and NICMOS images.
However, we have failed to identify a candidate from the USNO-B catalog.
The probability that a field star was responsible for the collision
in the past few thousand years
is very low, $\sim 10^{-6}$.  If the star responsible for
the encounter originated from a region (not yet identified)
with a higher stellar
density located within a few arcminutes of HD 100546 then
the probability of the encounter might be higher.
Here we have discussed three scenarios which could account for
the observed spiral structure in the disk of HD 100546.  Only
an extremely cold and thin disk could be unstable to the formation
of spiral density waves.  For the reasons discussed above, bound perturbers
also provide unlikely explanations.  This has left only the possibility
that a recent stellar encounter caused the spiral structure, however
this scenario is highly improbable and the star
responsible has not yet been identified. This implies
that the observed spiral structure currently lacks a good explanation.
A recent studies of the infrared spectral energy distribution  of HD100546
suggests that it's outer edge is flared \citep{dullemond,meeus}.  
Consequently structure observed in the visible bands could 
be due variations in the vertical structure.  Future scenarios
for HD100546's disk
might consider flared disk models that exhibit variations
in both vertical and azimuthal  structure.
Promising possibilities include transient spiral structure excited
by planetesimal collisions (e.g., \citealt{kenyon04}).

The division between bound perturbers and flybys is important.
While both excite spiral arms, a single flyby will not truncate
the disk. However an eccentric bound perturber, because it makes 
multiple close passages is likely to truncate the disk.   
Notably HD 141569A is an 
example of the second case but previous studies
have proposed that HD 100546 could be an 
example of the first case or a flyby. HD 100546 has a large disk 
extending out to 500AU, past the location of the spiral structure. 
HD 100546 also has 
an envelope that extends out to 1000AU \citep{grady}.
We expect that the process of disk truncation depends on
the velocity of the perturber.  If the perturber is moving fast,
then an impulse approximation is appropriate and disk particles
nearer the point of closest approach are more strongly affected than
those on the opposite side, particularly if the object is
low mass \citep{pfalzner}.  In contrast, the repeated encounters of 
the binary HD 141569B,C enforce a sharp outer disk boundary.

We note that the simulations presented here suffered from various restrictions.
They were 2-dimensional making it impossible to investigate the 
3-dimensional structure of these disks.  They suffered from artifacts
caused by the boundary conditions.  
Future studies could be 
carried out in 3 dimensions.  Larger grids could also be used,
lessening the artifacts caused by the boundary.
3D simulations may provide a better match to the outer morphology
of HD 141569A's disk, and make it possible to more tightly
constrain its disk thickness and companion's orbit.
In the future we also hope to simulate the effect of bound perturbers
and flybys on disks containing planets, in search of explanations
for the eccentric planets proposed to explain the structure of 
Vega's, HR 4796A's and Epsilon Eridani's disks 
\citep{wilner,wyatt,quillen_thorndike}.
The simulations carried out here were done using gaseous disks,
however these disks may contain massive planetesimals.    While
our flyby hydro simulations exhibit similar morphology to the collisionless 
simulations of \citet{pfalzner},
future work which investigates
the dynamics of disks containing both planetesimals that 
can collide and gas may be able to place constraints
on the constituents of these disks from their morphology.
Higher angular resolution, and deeper imaging spanning
additional wavelengths will also provide more constraints on
the dynamical processes affecting these disks.

Since most stars are born in stellar clusters and are binaries, 
encounters may strongly influence forming planetary systems.  
We find here that the binary companion of HD 141569A has likely truncated
and compressed its disk.  Future close encounters will also excite
spiral density waves in the disk.  This type of excitation,
taking place on timescales of thousands of years, could contribute
to periodic episodes of high accretion in gaseous disks or high 
dust production in planetesimal disks.
We note that flybys excite structure for only short periods of time,
so future studies can search for and identify objects 
capable of perturbing young stellar disks. 
Once the object responsible for exiting the spiral
structure is identified, the nature of the encounter can
be explored more quantitatively, and we expect that many
of the free parameters affecting our simulations (such as
orbit, disk thickness, boundary conditions and disk
composition) can be better constrained.  

\acknowledgments
We thank Mike Jura, Dan Watson, Pawel Artymowicz 
and Carol Grady for helpful discussions and comments.
We thank F.~Masset for his help and for providing the code.
Support for this work was provided by NSF grants AST-0406823, AST-9702484,
AST-0098442, 
DOE grant DE-FG02-00ER54600,
and the Laboratory for Laser Energetics.
This material is based upon work supported by the National Aeronautics
and Space Administration under Grant No. NNG04GM12G
issued through the Origins of Solar Systems Program and 
grant NAG5-8428, 
This research was supported in part
by the National Science Foundation to the KITP
under Grant No. PHY99-07949.

\clearpage

\smallskip

\clearpage

\begin{deluxetable}{lcll}
\tablecaption{Numerical simulations with exterior bound planets or stars\label{tab:erun}}
\tablewidth{0pt}
\tablehead{
\colhead{\#} &
\colhead{$q/10^{-3}$} &
\colhead{$e$} &
\colhead{$h/r$}  
}
\startdata
A & 1  & 0.0  &0.04 \\ 
B & 3  & 0.0  &0.04  \\ 
C & 3  & 0.25 &0.04  \\ 
D & 3  & 0.5  &0.04  \\ 
E & 5  & 0.0  &0.04  \\ 
F & 5  & 0.25 &0.04  \\ 
G & 5  & 0.5  &0.04  \\ 
H & 10 & 0.0  &0.04  \\ 
I & 10 & 0.25 &0.04  \\ 
J & 10 & 0.5  &0.04  \\ 
K & 5  & 0.25 &0.06  \\ 
L & 5  & 0.25 &0.10  \\ 
M &200 & 0.2  &0.04  \\ 
N &200 & 0.4  &0.04  \\ 
O &200 & 0.6  &0.04  \\ 
P &200 & 0.4  &0.10  \\ 
\enddata
\tablecomments{
The parameter $q$ is the ratio of the perturber's mass 
to the central stellar mass,  
$e$ is the planet's eccentricity, $h/r$ is the disk aspect ratio (setting
the sound speed).  For these simulations 
the disk Reynolds number ${\cal R} = 10^6$.
For simulations \#A--\#L,
the planet's semi-major axis is at 1.2 times $R_{max}$, 
the disk outer edge.
For simulations \#M--\#P the position of periapse $a(1-e)$ 
is fixed at 1.1 times $R_{max}$.
}
\end{deluxetable}

\begin{deluxetable}{llll}
\tablecaption{Numerical simulations 
with stellar parabolic encounters \label{tab:frun}}
\tablewidth{0pt}
\tablehead{
\colhead{\#} &
\colhead{$q$} &
\colhead{$R_a$} &
\colhead{$h/r$}  
}
\startdata
R & 0.3 & 0.5  &0.04  \\ 
S & 0.1 & 0.5  &0.04  \\
T & 0.05& 0.5  &0.04  \\
U & 0.1 & 0.5  &0.10  \\
\enddata
\tablecomments{
Encounters are parabolic with initial starting point 1.8 times the disk outer
radius. The parameter $R_a$ is the radius of closest 
approach times the disk outer radius.
}
\end{deluxetable}

\end{document}